 \documentclass[preprint,review,12pt]{elsarticle}

\usepackage{amssymb}
\usepackage{amsmath}

 \usepackage{lineno}

\journal{Nuclear Inst. and Methods in Physics Research, A}

\begin{document}
\begin{frontmatter}



\title{A magnetic spectrometer to measure electron bunches accelerated at AWAKE}


\author[cern]{J.~Bauche}
\author[cern,ctu]{B.~Biskup}
\author[ucl]{M.~Cascella}
\author[ucl]{J.~Chappell}
\author[cern]{N.~Chritin}
\author[ucl]{D.~Cooke}
\author[ucl]{L.~Deacon}
\author[cern]{Q.~Deliege}
\author[cern]{I.~Gorgisyan}
\author[cern]{J.~Hansen}
\author[ucl]{S.~Jolly}
\author[ucl]{F.~Keeble\corref{cor}}
\cortext[cor]{Corresponding author at: UCL, London, UK}
\ead{fearghus.keeble.11@ucl.ac.uk}
\author[eso]{P. La Penna}
\author[cern]{S.~Mazzoni}
\author[cern]{D.~Medina~Godoy}
\author[cern]{A.~Petrenko}
\author[eso]{M.~Quattri}
\author[cern]{T.~Schneider}
\author[ucl]{P.~Sherwood}
\author[cern]{A.~Vorozhtsov}
\author[ucl]{M.~Wing}

\address[cern]{CERN, Geneva, Switzerland}
\address[ctu]{Czech Technical University, Prague, Czech Republic}
\address[ucl]{UCL, London, UK}
\address[eso]{ESO, Munich, Germany}

\begin{abstract}
A magnetic spectrometer has been developed for the AWAKE experiment at CERN in order to measure the energy distribution of bunches of electrons accelerated in wakefields generated by proton bunches in plasma. AWAKE is a proof-of-principle experiment for proton-driven plasma wakefield acceleration, using proton bunches from the SPS. Electron bunches are accelerated to $\mathcal{O}$(1\;GeV) in a rubidium plasma cell and then separated from the proton bunches via a dipole magnet. The dipole magnet also induces an energy-dependent spatial horizontal spread on the electron bunch which then impacts on a scintillator screen. The scintillation photons emitted are transported via three highly-reflective mirrors to an intensified CCD camera, housed in a dark room, which passes the images to the CERN controls system for storage and further analysis. Given the known magnetic field and determination of the efficiencies of the system, the spatial spread of the scintillation photons can be converted to an electron energy distribution. A lamp attached on a rail in front of the scintillator is used to calibrate the optical system, with calibration of the scintillator screen's response to electrons carried out at the CLEAR facility at CERN. In this article, the design of the AWAKE spectrometer is presented, along with the calibrations carried out and expected performance such that the energy distribution of accelerated electrons can be measured.
\end{abstract}


\end{frontmatter}



\section{Introduction}
\label{sec:introduction}

The Advanced Wakefield (AWAKE) experiment at CERN is a proof-of-principle experiment demonstrating plasma wakefield acceleration using a proton drive beam for the first time~\cite{Assmann:2014hva,Caldwell:2015rkk,Gschwendtner:2015rni,Muggli:2017rkx}. Proton bunches from the CERN SPS accelerator are injected into a rubidium (Rb) vapour and co-propagate with an intense laser pulse which creates the plasma and seeds the modulation of the proton bunch into microbunches~\cite{PhysRevLett.122.054802,PhysRevLett.122.054801}. These microbunches induce strong resonant wakefields which are sampled by an externally-injected electron bunch, which is accelerated to high energy.\par
A magnetic spectrometer has been installed downstream of the plasma cell in order to measure the energy distribution of the accelerated electron bunch. The spectrometer has been designed to fulfil the following requirements:
\begin{itemize}
\item Separate the accelerated electrons from the drive bunch protons.
\item Introduce a spatial distribution into the accelerated bunch that is a function of energy.
\item Measure the spatial intensity distribution of the accelerated electrons to allow the mean energy, energy spread and bunch charge to be calculated.
\item Provide sufficient acceptance to prevent significant loss of accelerated electrons before the energy measurement.
\item Provide sufficient dynamic range to allow measurement of a range of electron energies from 0--5 GeV.
\item Measure the energy profile of the electron bunch with sufficient resolution to demonstrate proton-driven plasma wakefield acceleration of witness bunch electrons.
\end{itemize}
The AWAKE electron spectrometer has been used recently to measure acceleration of electrons to GeV energies in the first demonstration of proton-driven plasma wakefield acceleration~\cite{Adli:2018}. The evolution of the spectrometer's design has been discussed previously~\cite{Deacon:2141860,Keeble:IPAC2018-THPML118}. Here, we present the final design and full calibration of the system.

\subsection{Overview}
\label{subsec:overview}

\begin{figure}[t]
\includegraphics[width=\columnwidth]{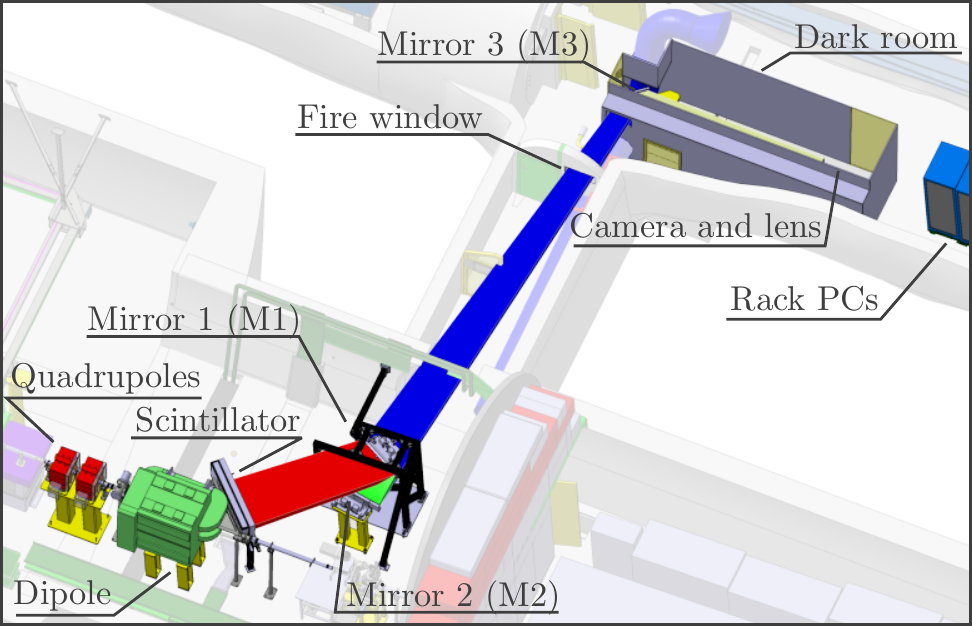}
\caption{The electron spectrometer at AWAKE. The path of the scintillator photons which reach the camera is shown with coloured blocks: the red block shows the path from the scintillator to the first mirror (M1), the green block shows the path from M1 to the second mirror (M2), the blue block shows the path from M2 through the fire safety window to the third mirror (M3) which is within the spectrometer dark room and the yellow block shows the path inside the dark room from M3 to the lens and camera. Close to the dark room are the rack PCs used for data acquisition and control of the camera.}
\label{fig:cad}
\end{figure}

The layout of the spectrometer within the AWAKE tunnel is shown in Figure~\ref{fig:cad}. The magnetic part of the spectrometer system begins approximately 4.5\,m downstream of the plasma cell exit and consists of two quadrupoles followed by a C-shaped dipole magnet. Inside the dipole magnet the AWAKE beamline expands into a large triangular vacuum chamber, terminated on one side by a thin window which allows high energy electrons to pass through. Attached to the exterior surface of the window is a scintillating phosphor screen which emits photons when particles deposit energy in it. The scintillator photons are transported, via a series of large mirrors, to a focusing lens and CCD camera in an adjacent tunnel.

\section{Components}
\subsection{Magnets}
\begin{figure}[t]
\includegraphics[width=\columnwidth]{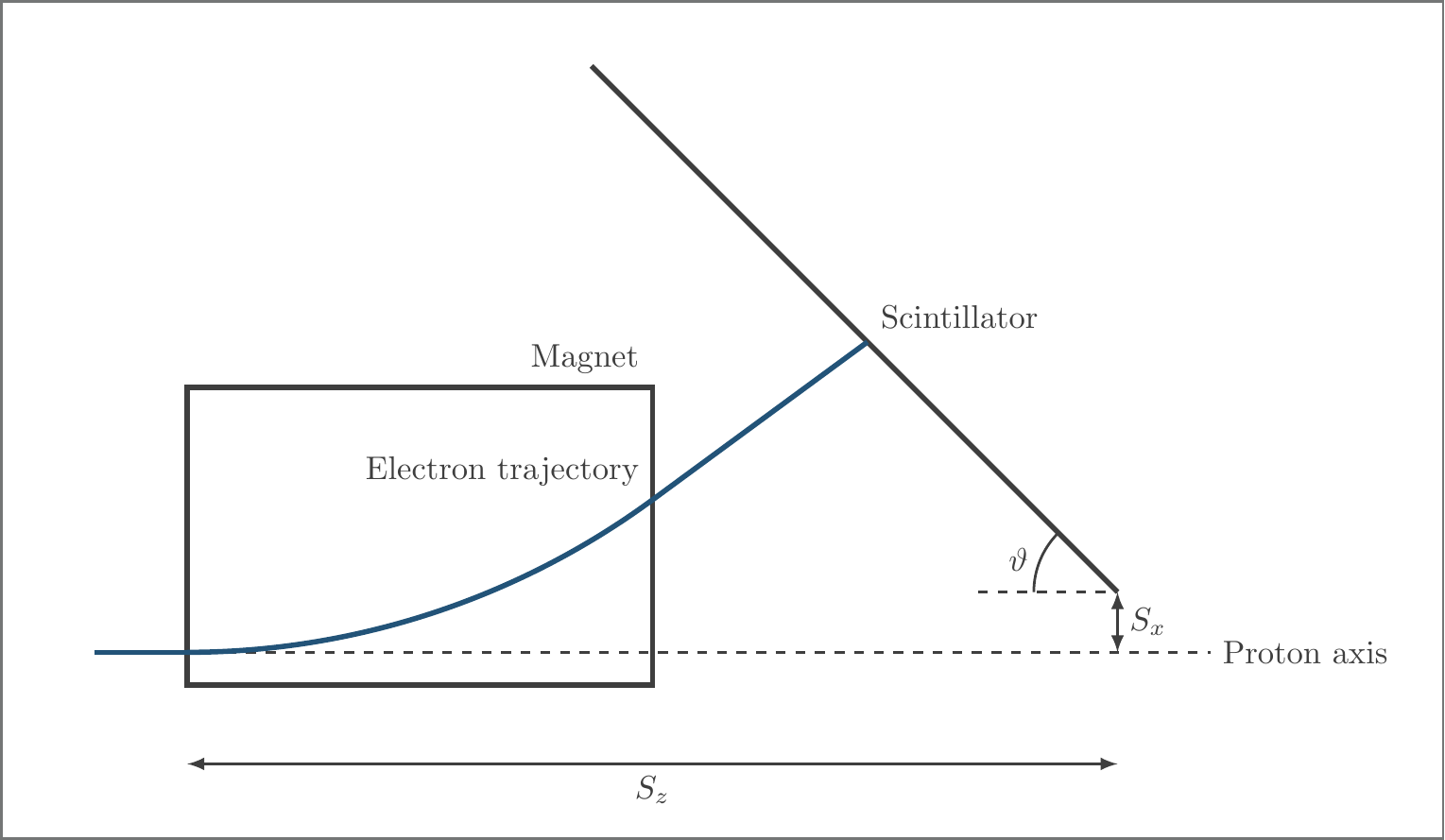}
\caption{A minimal schematic defining each of the quantities listed in Table~\ref{tab:uncerts}. An example of an electron trajectory, propagating from left to right, is shown in blue.}
\label{fig:diag}
\end{figure}

The spectrometer dipole is an electromagnet which can be stably operated between input currents of 18\,A and 650\,A, corresponding to approximate integrated magnetic fields of 0.065\,T\,m and 1.545\,T\,m respectively. The length of the magnet's iron is 1\,m in the direction parallel to the beamline and 0.32\,m in the transverse direction. To reduce the impact of fringe fields on the electrons while maintaining a large integrated field the magnet is offset in the transverse direction such that electrons have 0.285\,m of iron in the direction in which they are bent. At the lowest magnet setting, electrons at the injection energy of approximately 18\,MeV can be measured by the spectrometer and at the highest setting, electrons with energies up to 8.5\,GeV can be measured. The strength of the magnetic field is varied by changing the current through the magnet's coils. The field has been mapped for a number of these currents and finite element analysis (FEA) simulations have been performed to infer field maps for other current settings. With these field maps the position--energy conversion function for the spectrometer can be specified using only three additional parameters. These parameters are displayed in Figure~\ref{fig:diag} and are summarised in Table~\ref{tab:uncerts}. The measurements come from a combination of a dedicated survey and measurements of the proton bunch's position. \par
\begin{figure}[t]
\includegraphics[width=\columnwidth]{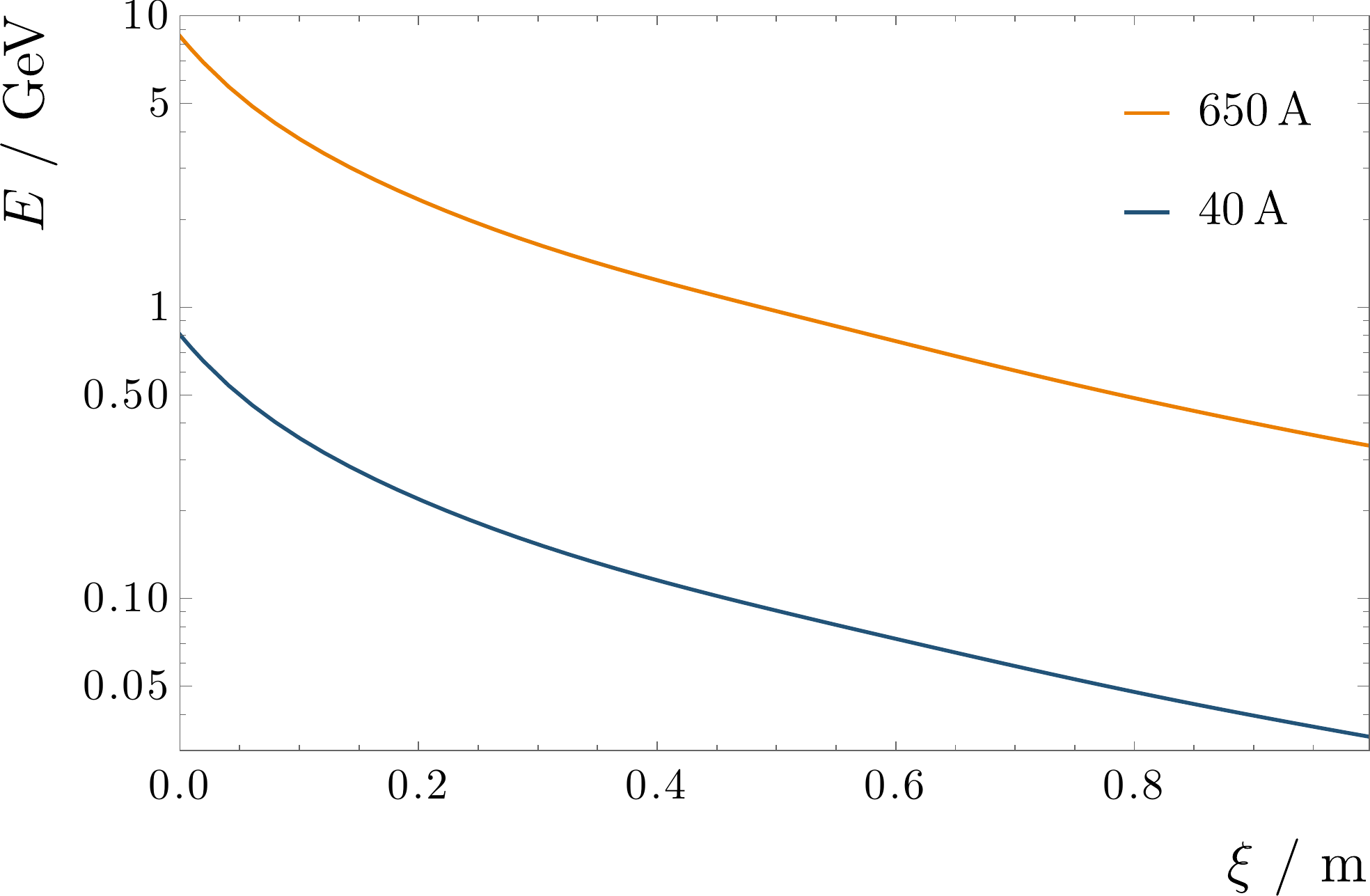}
\caption{The position--energy ($\xi$--$E$) relationship at dipole magnet settings of 40\,A and 650\,A, simulated using BDSIM.}
\label{fig:conversion_funcs}
\end{figure}

With the above measurements, the position--energy conversion function can be simulated using BDSIM~\cite{Nevay:2018zhp}. This simulation can be compared to an analytic solution under the assumption of a uniform magnetic field and the results are found to match to within 2\% at any given point on the scintillator. The uncertainty in the conversion function arising from uncertainties in the measured values was also estimated in these simulations. However, a 1\% overall uncertainty in the magnetic field map, determined by comparing the available measured values to those simulated by FEA, dominates over the uncertainties shown in Table~\ref{tab:uncerts}. Examples of the position--energy function using two of the field maps for input currents of 40\,A and 650\,A are shown in Figure~\ref{fig:conversion_funcs}.  At 40\,A the energy range available is approximately 30--800\,MeV and at 650\,A it is 300--8500\,MeV. The relationship between the position and energy is non-linear, changing slowly at the lower energy end of the scintillator and rapidly at the high energy end. This has important implications for the energy resolution, as discussed in Section~\ref{sec:op_cal}.
\par
The spectrometer's quadrupoles have an iron length of 0.285\,m and a peak magnetic field gradient of 18.1\,T\,m$^{-1}$ at a current of 362\,A. At this setting the quadrupoles are maximally focusing for a beam of approximately 1.3\,GeV. Because the quadrupoles are separated by about 0.2\,m, they must be offset in strength by approximately 6\% in order to both focus onto the plane of the scintillator. However, the electron's path length from the quadrupoles to the scintillator varies depending on which part of the scintillator they are incident upon and, hence, their energy. This variation in the path length of 0.35\,m from the high energy end to the low energy end means that the quadrupoles cannot be offset with respect to each other to focus perfectly at the screen for all energies and the offset of 6\% is a compromise to provide reasonable focusing across the whole screen.

\begin{table}
\centering
\caption{Measured values for each of the parameters defined in Figure~\ref{fig:diag}.}
\label{tab:uncerts}
\begin{tabular}{ |c|c| }
\hline
\textbf{Parameter} & \textbf{Value} \\
\hline
$S_{z}$         &         $1.676\pm0.001$\,m \\ 
\hline
$S_{x}$         &         $0.0620\pm0.0005$\,m \\ 
\hline
$\vartheta$         &         $44.80\pm0.01^{\circ}$ \\
\hline
\end{tabular}
\end{table}

\subsection{Camera and optical line}
The camera used to image the spectrometer's scintillator is an Andor iStar 340T, an intensified camera with a 2048\,$\times$\,512 pixel CCD, often used in low-light and spectroscopy applications. The camera is triggered approximately 100\,ms before proton extraction to AWAKE occurs and is delayed internally using the camera's digital delay generator which controls when the intensifier is gated to amplify the light. The camera is controlled remotely using a bespoke FESA~\cite{arruat2007front} class which is interfaced to the camera using Andor's SDK. This class is also responsible for data readout and interfaces to AWAKE's data acquisition and logging systems. To reduce readout noise during operation the camera is cooled to $-30^{\circ}\mathrm{C}$ using an in-built Peltier device with a heat sink cooled by a closed-circuit liquid cooling system circulating a 2:1 mixture of distilled water and ethylene glycol at $12^{\circ}\mathrm{C}$. \par
A unique challenge for the spectrometer is the high level of radiation in the AWAKE tunnel, generated by the proton drive bunch. This radiation necessitates placing the spectrometer's camera far away from the beamline to reduce the background noise and protect it from radiation damage. This requires a specially designed optical line consisting of a long focal length lens and three mirrors. \par
The optical distance between the camera and the scintillator is 17\,m. To ensure sufficient light capture and resolution a long focal length, low f-number lens is used: a Nikon AF-S NIKKOR 400 mm $f$/2.8E FL ED VR. The front of the lens is fitted with a 550$\pm$50\,nm filter of the same diameter. This filter reduces the ambient background due to lights in the experimental area. The parameters for this lens and the dimensions of the scintillator were used as inputs to a Zemax OpticStudio simulation to define the required dimensions for the optical line mirrors. These dimensions are summarised in Table~\ref{tab:mirrors}, which shows both the required size of the mirror (clear aperture) as defined by Zemax OpticStudio and the physical size used in the experiment.  The image intensifier in the CCD camera limits the active pixels in the horizontal axis to 1850. For the 0.997\,m wide scintillator this gives 0.54\,mm\,px$^{-1}$. Imaging a resolution target directly from 17\,m with the camera shows that the resolution limit is approximately 1.5\,mm with the limiting factor likely being the intensifier in the camera. To maintain this resolution, the Zemax OpticStudio simulation of the line shows that the mirrors must be optical grade; they must have $\lambda/2$ flatness over any 100\,mm area. Additionally, the scratch-dig of the mirrors must not exceed 80/50. The mirrors are made from BK7 glass which is polished to achieve the desired flatness and scratch-dig. This polishing process generates a considerable amount of heat and, given the thermal expansion of the BK7 and the required surface properties, this necessitates a relatively thick piece of glass. As such, each mirror has a thickness of 40\,mm. This thicker glass is also minimally affected by gravitational bending. This is particularly important for M1 which hangs facing downwards, with the mirror held in place by the three adjustment screws.\par

\begin{table}
   \centering
   \caption{Mirror dimensions and clear apertures. M1 is the mirror closest to the scintillator.}
   \begin{tabular}{|l|c|c|}
       \hline
       \textbf{Mirror} & \textbf{Width / mm}                      & \textbf{Height / mm} \\
       \hline
          M1 full         & 926.0            & 150.0       \\ 
          \hline
          M1 clear aperture       & 898.2           & 121.5       \\ 
          \hline
          M2 full         & 926.0            & 150.0       \\ 
          \hline
          M2 clear aperture       & 819.5           & 126.4       \\ 
          \hline
          M3 full         & 524.0            & 160.0       \\ 
          \hline
          M3 clear aperture       & 504.6           & 140.5       \\ 
       \hline
   \end{tabular}
   \label{tab:mirrors}
\end{table}

The polished glass has a protected aluminium coating, with three layers of material designed to ensure high reflectance, uniformity and ease of use. The first is a 10\,nm chromium layer to ensure adhesion of the coating to the glass. The second layer is 100\,nm of aluminium which was selected because of its good reflectance around the 545\,nm peak of the scintillator emission. The final layer is 185\,nm of quartz, to further enhance reflectance and to prevent the oxidation of the aluminium layer. The thickness of the quartz layer has been adjusted using a combination of simulation and testing such that the mirrors provide their most uniform reflectance around the wavelength of emission of the scintillator.  The mirrors were coated by evaporation; via an electron gun for the quartz and chromium layers and thermally for the aluminium layer. Each mirror has a reflectance of approximately 92\% around the emission peak of the scintillator.\par
Due to the mirror's large size, bespoke mounts have been designed. The tip and tilt of the mirrors may be adjusted in these mounts using three screws and the mounts themselves can be further adjusted by additional screws. The mounts have been designed to hold the mirrors securely to minimise the need to realign the system. Another key feature of the mounts is that they are sturdy, such that vibrations from the floor are significantly damped. These vibrations can lead to movement of the mirrors which blurs the images recorded by the camera. The mirror most affected by these movements is M1 due to its less rigid base and the fact that the mirror hangs from the mount rather than sitting on it.  A harmonic FEA modelling has been performed using ANSYS to determine the transfer functions of the mount for M1. These transfer functions in combination with the measured power spectral density from the mount locations can be used to determine the displacement of the centre of gravity of the mirror.  Inputting these displacements into a model of the optical line in Zemax OpticStudio shows that there is a negligible effect on the image. Furthermore, the analysis shows that the vibrations do not cause movements of the mirrors at frequencies higher than 1\,Hz. This is important since the exposure time of the camera is typically $\mathcal{O}(100\,\mu$s) and, as such, movements below approximately 1\,kHz would cause no appreciable blurring of the images.\par
The only other optical component in the line is a $550\times200\times3$\,mm$^{3}$ BK7 window between mirrors M2 and M3. This window is located in a door and its purpose is to maintain the fire rating of the door while minimally affecting the light passing through it. The window is coated with a broadband anti-reflective coating giving an overall transmittance of greater than $99.0\%$ in the wavelength range $550\pm50$\,nm.

\subsection{Vacuum chamber and window}
\begin{figure}[t]
\includegraphics[width=\columnwidth]{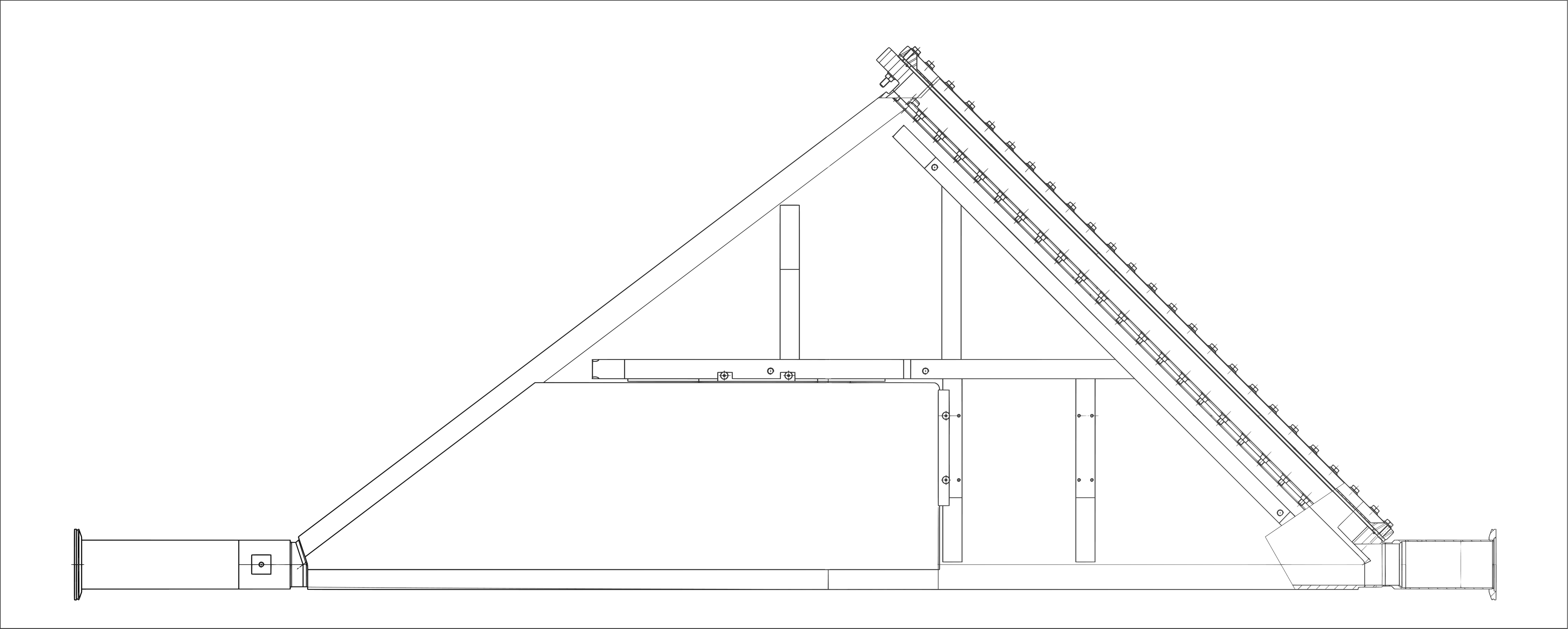}
\caption{Schematic of the spectrometer vacuum chamber. Electrons enter from the left hand side and are spread out by the magnetic field of the magnet. The vacuum window comprises the upper right side of the triangular chamber and the scintillator is fixed to its exterior surface. The high energy protons propagate through the chamber and exit through the beam pipe on the right hand side, continuing to a beam dump.}
\label{fig:vc}
\end{figure}

The large spectrometer magnet necessitates the use of a bespoke vacuum chamber, a schematic of which is shown in Figure~\ref{fig:vc}. A large portion of the vacuum chamber is positioned within the aperture of the magnet and, as such, its full height is restricted to 80\,mm. To minimise the loss of accelerated electrons this aperture is kept as clear as possible. This means that the portion of the chamber inside the magnet has no stiffeners. The chamber has a 6\;mm thick stainless steel wall, to prevent buckling. The centre of the chamber is attached to the magnet for the same reason.
\par
The most delicate part of the vacuum chamber is the window through which the electrons pass. To minimise the scattering and loss of these electrons the window is composed of a 2\;mm thick aluminium alloy. To avoid welding, which could create weak points, the whole window assembly is produced from one solid sheet of aluminium 6082-T6. The aluminium grain size is approximately 10--20\;$\mu$m meaning that there is a minimum of 100 grains across the thickness of the window. This relatively large number of grains is sufficient to ensure that the window is leak-tight. The window is 62\,mm high and 997\,mm wide with the aluminium rounded into a semicircle at each end to avoid corners which could create weak points.
\par
Electrons passing through the vacuum window scatter, losing energy and producing secondary particles in the process. The window is manufactured with a tolerance of 0.05\;mm and simulations show that an increase or decrease of this amount induces a change of only 5\;$\mu$m in the spatial spread of a 20\;MeV electron bunch after the window.

\subsection{Scintillator}
The scintillator chosen for the spectrometer is a DRZ-High screen, a terbium doped gadolinium oxysulfide (Gd$_{2}$O$_{2}$S:Tb) scintillator manufactured by Mitsubishi. The scintillator has a thickness of 507\,$\mu$m and has been cut to fit the vacuum window in the shape described above. The scintillator is attached to the exterior surface of the vacuum window using a 200\,$\mu$m thick double sided adhesive. This minimises the gap between the two components and, hence, the spread of the electron beam. The majority of the scintillator's emission is sharply peaked around 545\,nm and the response to incident radiation is linear for the charge densities present at AWAKE.

\section{Calibration and simulation}
\subsection{Optical line}
\label{sec:op_cal}
\begin{figure}[t]
\includegraphics[width=\columnwidth]{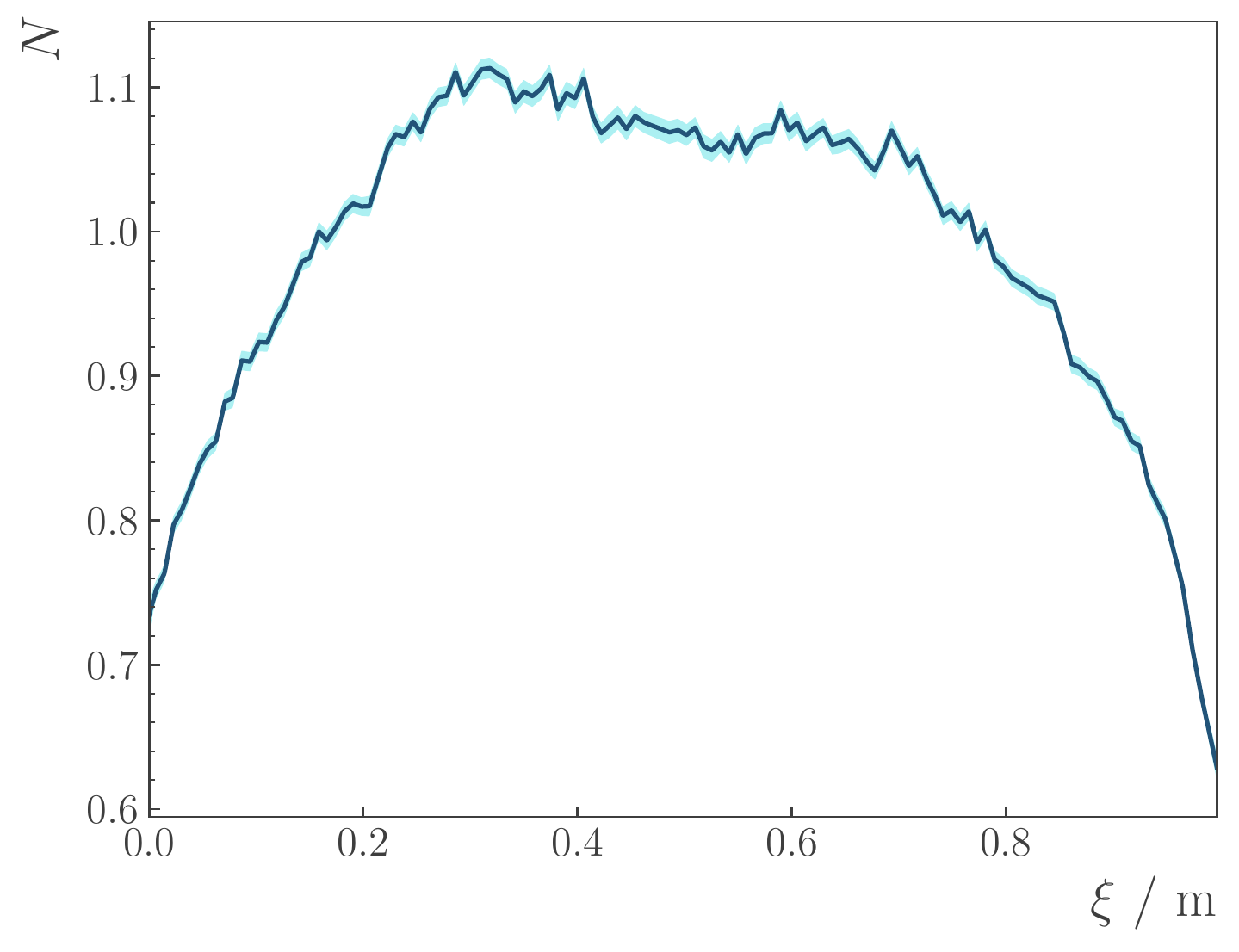}
\caption{Camera response to a constant light source scanned across the surface of the scintillator. The curve has been normalised to 1 at the point $\xi=0.84$\,m.}
\label{fig:vignette}
\end{figure}

\begin{figure}[t]
\includegraphics[width=\columnwidth]{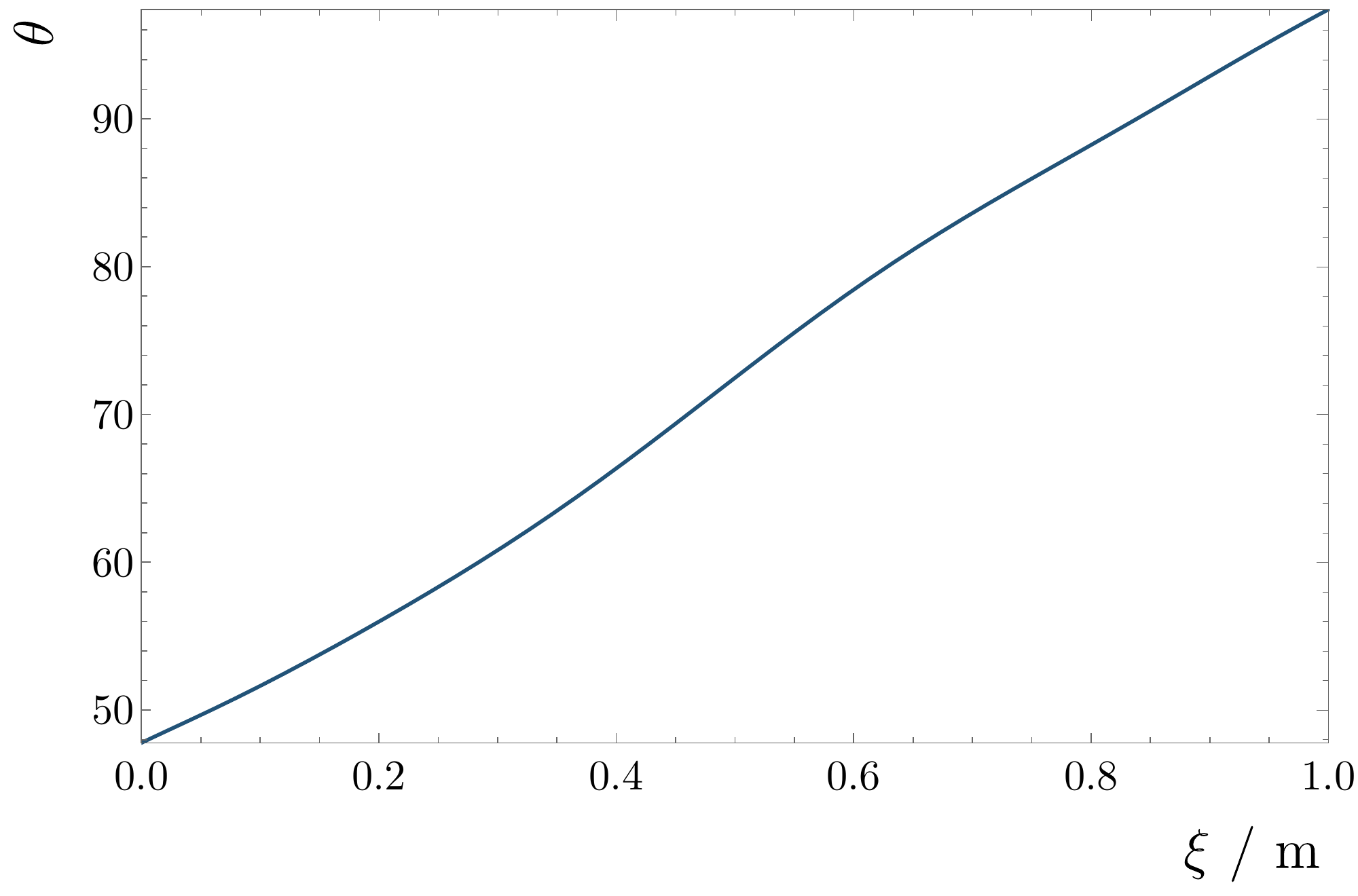}
\caption{Electron incident angle on the scintillating screen as a function of screen position as simulated using BDSIM. This curve is approximately independent of the magnet setting though some small variations do occur.}
\label{fig:angle}
\end{figure}

Not all the light produced by the scintillator is captured, due to the angular emission profile and the finite size of the spectrometer's optics. This induces a position dependence in the light captured by the camera, which must be corrected for. This correction factor is measured by imaging a constant light source as it is scanned across the surface of the scintillator. This light source is a diffuse emitter of light peaked at 545\,nm, mimicking the scintillator.  This scan produces a curve which allows the scintillator emission to be normalised relative to a given point. In Figure~\ref{fig:vignette} the curve has been normalised such that the emission measured at $\xi=0.84$\,m is 1. This is the point at which electrons are normally incident on the scintillator, as shown in Figure~\ref{fig:angle} which gives the incident electron angle $\theta$ for each $\xi$.
\par
\begin{figure}[t]
\includegraphics[width=\columnwidth]{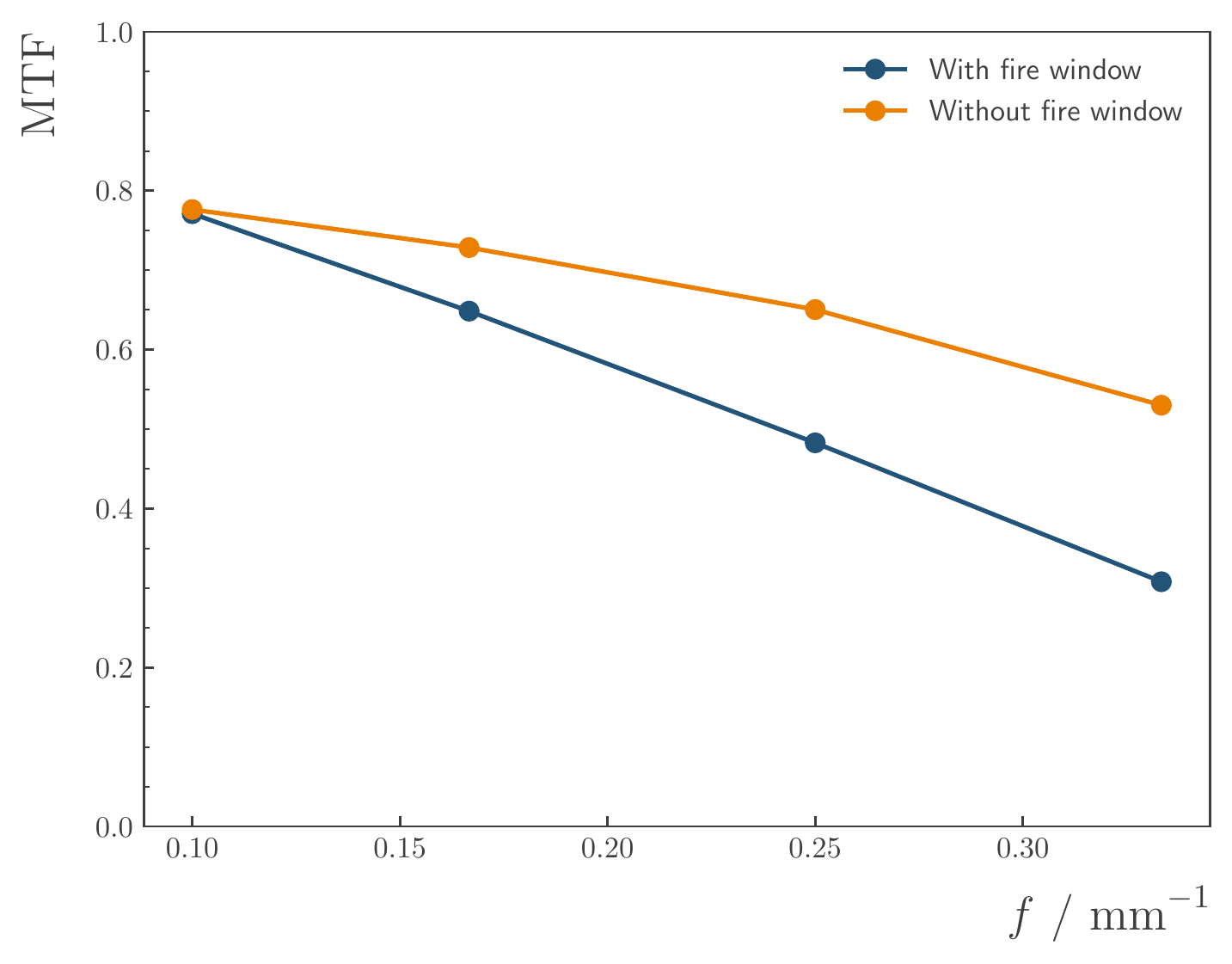}
\caption{Modulation transfer function (MTF) of the full optical system measured using a resolution target. The four sets of bars in the target are spaced with frequency $f$. Results are shown for the system with and without the fire safety window present.}
\label{fig:mtf}
\end{figure}

The optical resolution of the system is also determined using the light source. A resolution target consisting of a number of black and clear bars of varying widths is fixed to the front of the light source and imaged.  From this, the modulation transfer function (MTF) of the system and, hence, the resolution, may be determined. The MTF for the optical system is shown in Figure~\ref{fig:mtf}. Without the fire safety window present the system performs as designed; the MTF is above 0.5 for a spatial frequency of 0.33\,mm$^{-1}$, corresponding to a resolution of 1.5\,mm. However, the inclusion of the 3\,mm thick fire safety window significantly affects the MTF at higher spatial frequencies, limiting the resolution to approximately 2\,mm.
\par

\begin{figure}[t]
\includegraphics[width=\columnwidth]{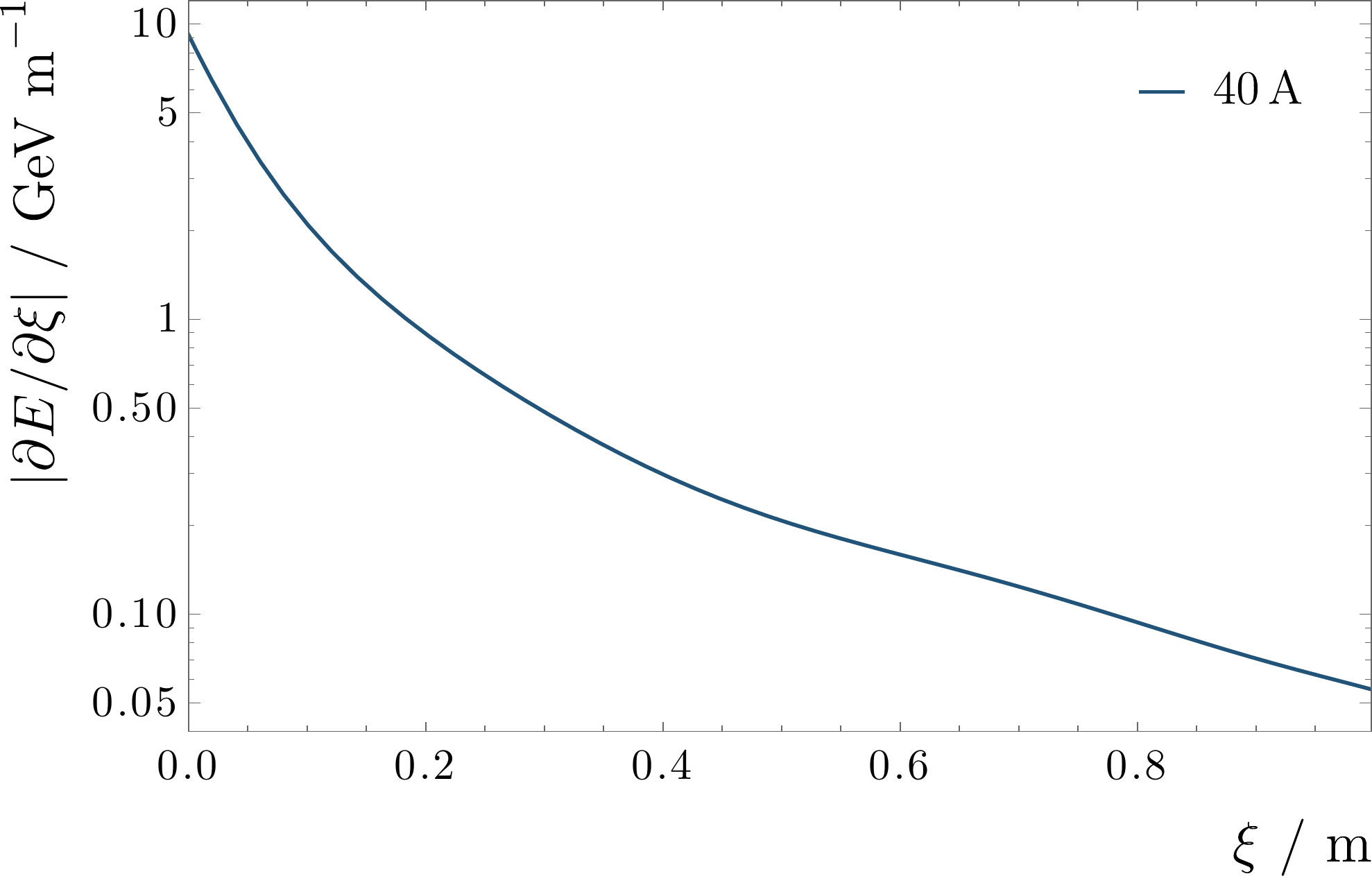}
\caption{Absolute energy derivative with respect to position across the screen for a 40\,A dipole setting. The curve peaks at the high energy end showing that measurements in this region are more sensitive to an uncertainty in the electron's $\xi$ position.}
\label{fig:deriv_40}
\end{figure}

This 2\,mm optical resolution limit restricts the energy resolution. This effect is particularly significant at high energies, as displayed in Figure~\ref{fig:deriv_40} which shows the derivative of the inferred energy with respect to $\xi$ across the scintillator for a 40\,A dipole current. At the high energy ($\sim$800\,MeV) end a 2\,mm uncertainty in $\xi$ corresponds to an energy uncertainty of approximately 19\,MeV, or 2.4\%, which is larger than the 1\% uncertainty arising from the magnetic field and dominates the overall uncertainty.

\par
\begin{figure}[t]
\includegraphics[width=\columnwidth]{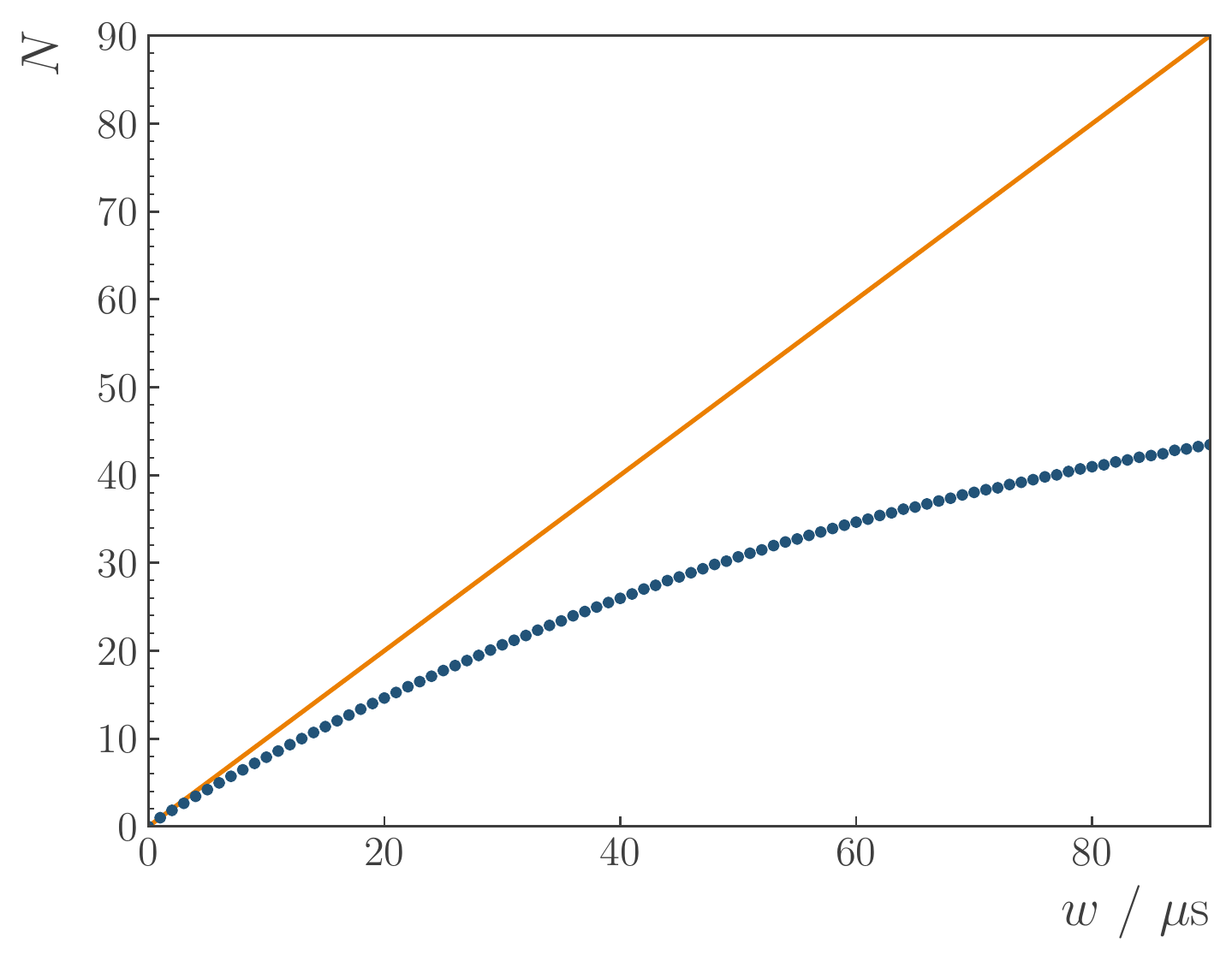}
\caption{Camera response to a constant light source for different gate widths $w$ (blue), normalised such that a 1\,$\mu$s gate gives a response of 1. The points may be compared to a linear response in the gate width (orange).}
\label{fig:width}
\end{figure}
\begin{figure}[t]
\includegraphics[width=\columnwidth]{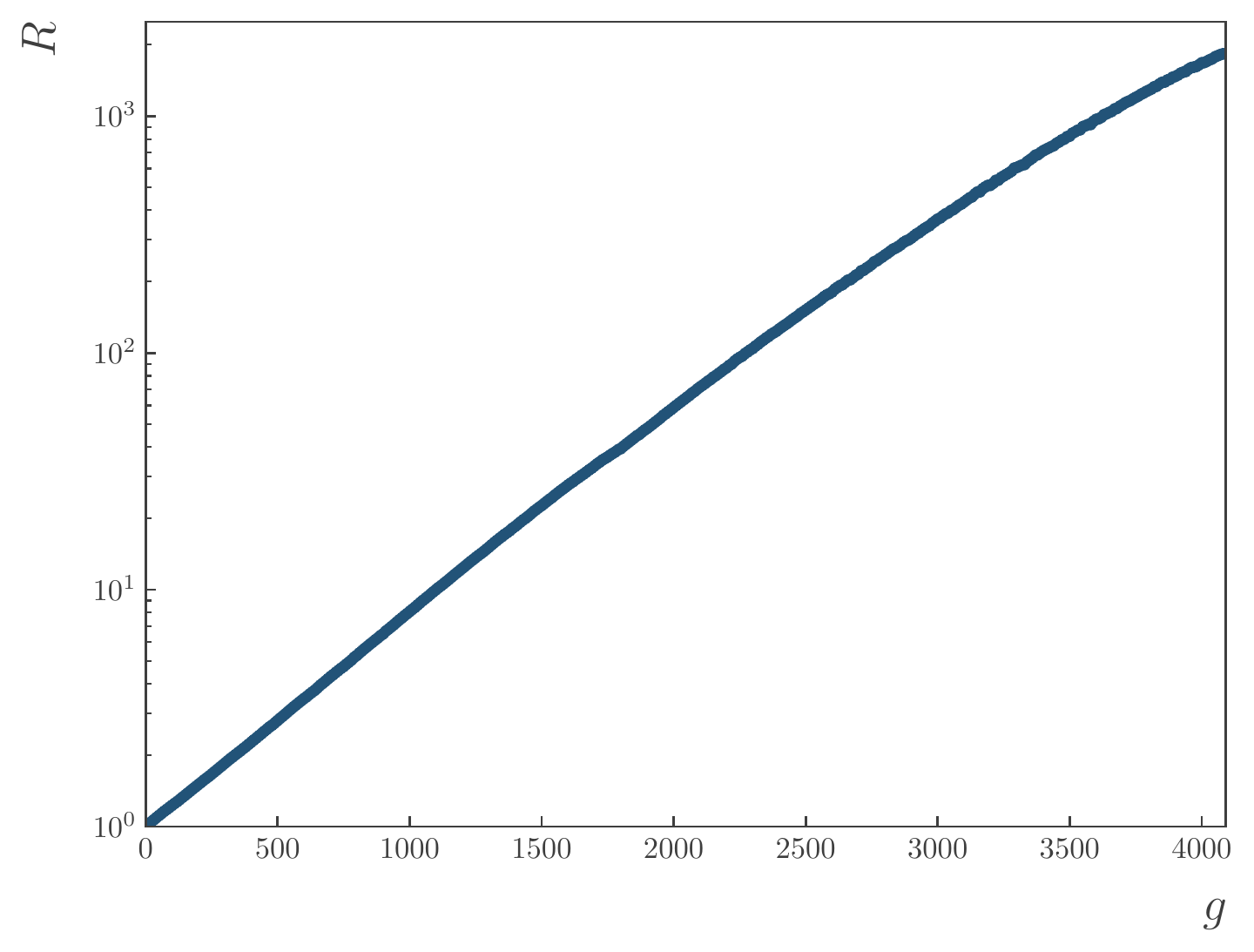}
\caption{Camera response $R$ to a constant light source for different gain values $g$ (blue), normalised such that a gain of 0 gives a response of 1.}
\label{fig:gain}
\end{figure}

Changing the camera's gain and gate width is necessary to prevent saturation of the image under different conditions. For example, standard running conditions at AWAKE require a 500\,$\mu$s gate with a gain of 3000, but this does not work for calibration because the lamp is brighter than a typical signal. As such, the correction between different settings has been measured using the lamp. Increasing the gate width does not result in a linear increase in signal, as shown in Figure~\ref{fig:width}, which shows a plot of the camera response to a constant light source for different gate widths $w$. The points represent the mean of several measurements which have had a $w=0$ exposure subtracted and have been normalised such that a 1\,$\mu$s gate gives a response of 1. The orange line represents a linear 1:1 response where doubling the gate width results in a doubling of the signal. The correction for the camera's gain is shown in Figure~\ref{fig:gain}, where each point again represents the mean of several background-subtracted exposures to a constant light source. The response is approximately exponential at low to intermediate gain values but deviates at higher gains.

\subsection{Scintillator}
\begin{figure}[t]
\includegraphics[width=\columnwidth]{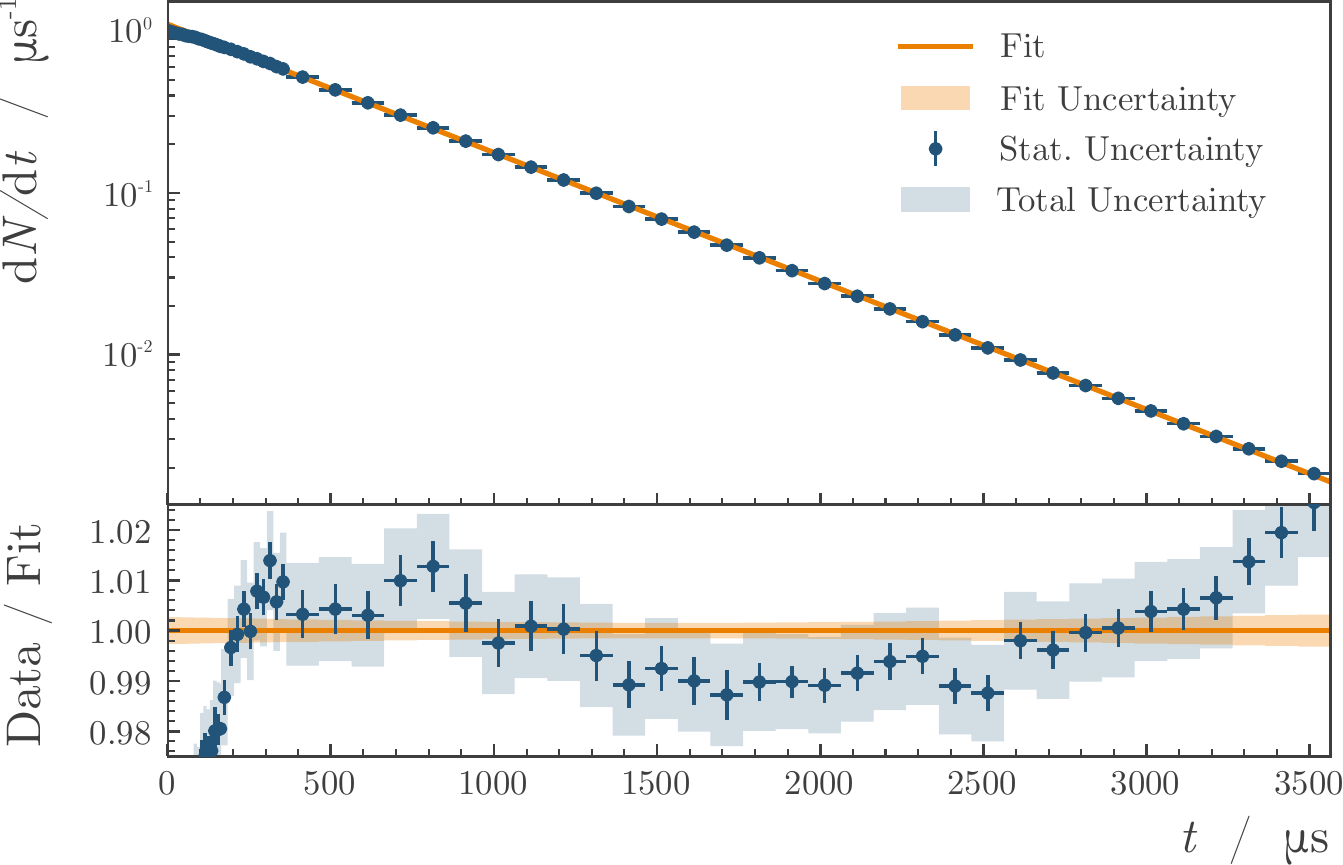}
\caption{Normalised response of the scintillator to radiation at different times after it passes. The radiation is incident at approximately 101.0363\,ms and the response reaches a maximum within 1\,$\mu$s of that. The response then decays, slowly at first and approximately exponentially after 200--250\,$\mu$s. The vertical bars are a combination of statistical and systematic uncertainties. The horizontal bars indicate the exposure time, not the timing uncertainty. The exponential fitted here has a half life of $379\pm1$\,$\mu$s.}
\label{fig:decay}
\end{figure}

When imaging the scintillator response with different gate widths an additional correction must be applied. This arises because the response of the scintillator to radiation varies over time, rising quickly to a maximum value and then decaying approximately exponentially with a given decay constant. This response has been measured using radiation generated by the proton beam at AWAKE. The radiation was generated by the proton beam interacting with a number of removable foils along the AWAKE beamline and the radiative flux is linearly proportional to the proton bunch charge, which varies but is measured each extraction. A scan was performed by increasing the delay before the camera is gated and taking a number of images at each setting. These images were background subtracted and then fit against the proton charge with a linear function with an intercept of 0. The fit coefficients for a series of delay settings are shown in Figure~\ref{fig:decay}, with an exponential function fit to the points with delays greater than 163.7\,$\mu$s. The axis is set such that the proton bunch passes at approximately 0\,$\mu$s and smaller gate widths have been used closer to this point to show the structure of the response. The data have been normalised relative to the first data point, which is centred on 1.2\,$\mu$s and has a gate width of 1\,$\mu$s.  As can be seen, the signal rises very quickly and then becomes well described by an exponential approximately 200\,$\mu$s after the radiation passes.  The error bars on each point come from a combination in quadrature of the statistical uncertainty and a systematic uncertainty arising from different experimental setups. The exponential fit over the full range returns a half life of $379\pm1$\,$\mu$s.\par
The long distance between the electron source and the spectrometer at AWAKE makes it difficult to propagate a bunch of well-known charge to the scintillator. Consequently the charge response of the scintillator has been measured at the CLEAR facility at CERN~\cite{GAMBA2018480}. A setup intended to mimic that present at AWAKE was used, with the scintillator attached to the vacuum window placed in the path of an electron bunch with an energy of approximately 150\,MeV.  The bunch was normally incident on the rear surface of the vacuum window with a spot size of $\mathcal{O}(1$\,mm). The charge of this bunch was scanned from the minimum available charge of approximately 2\,pC up to 35\,pC; a range intended to be representative of the expected accelerated bunch charge at AWAKE~\cite{Caldwell:2015rkk}. For each event the  charge was measured immediately before the bunch was incident on the vacuum window and the scintillator response was captured using the same Andor camera used at AWAKE. Due to the variable bunch charge for any given setting, a large number of images were taken at each point. The optical setup for the calibration was very different to that used at AWAKE. A smaller 105\,mm focal length lens was used and the camera was positioned at a distance of 3\,m from the scintillator facing orthogonal to the beamline, with the light reflected via a 5\,cm diameter silver mirror. The correction for the different optical setups is made using the same light source as used in the optical calibration, which mimics the emission of the scintillator. The wide charge range necessitated changing the camera gain and gate width settings for different points and these have been corrected for as described in the previous subsection. When the gate width is corrected the scintillator emission is also corrected using the half life measured in Figure~\ref{fig:decay}.
\par
\begin{figure}[t]
\includegraphics[width=\columnwidth]{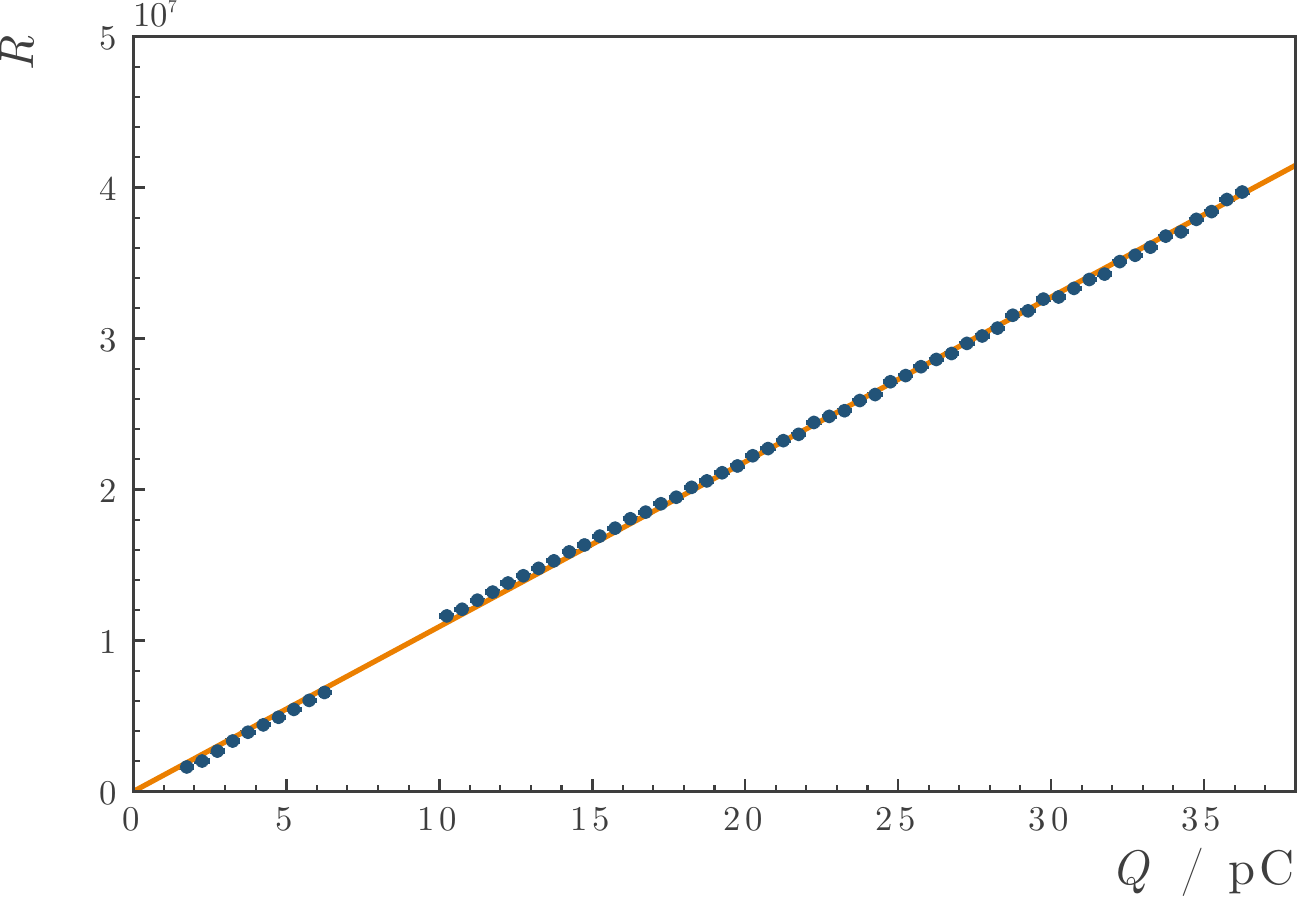}
\caption{Scintillator response to incident electron beam charge (blue) fitted with a linear function (orange). The data are binned by charge in 0.5\,pC bins.}
\label{fig:clear}
\end{figure}

 
The captured images were subtracted for three different backgrounds: the intrinsic camera background, the ambient background in the room and a particle background generated by radiation directly incident on the camera during events. The data are binned by charge with a bin width of 0.5\,pC, the approximate resolution of the charge measurement device.  The fit to the data is shown in Figure~\ref{fig:clear}, with a response of $1.09\pm0.02\times10^{6}$ CCD~counts~per~incident~pC. The values given here correspond to a gate width of 500\,$\mu$s and a gain of 250 measured from a delay of 200\,$\mu$s after the initial glow of the scintillator begins. Measurements of the scintillator at different points and for beam energies of 95 and 120\, MeV agree with this fit to within $1\sigma$. 
%

\section{Conclusion}
In conclusion, a magnetic spectrometer to measure accelerated electrons bunches at AWAKE has been designed. The spectrometer tackles the unique challenge of the high proton backgrounds present by removing the imaging device from the beamline area and transferring scintillation signals to it using an optical path comprised of metre-scale mirrors. The scintillator and the optical system have been sufficiently characterised in order to allow the spectrometer to achieve its goal of measuring the charge and energy of the accelerated electrons.

\section{Acknowledgements}
This work was supported by a Leverhulme Trust Research Project Grant RPG-2017-143 and by STFC, United Kingdom. We gratefully acknowledge F. Galleazzi for the model of the AWAKE tunnels used in Figure~\ref{fig:cad} and the operators of AWAKE, the SPS and the CLEAR facility for the provision of the proton and electron data used to produce Figure~\ref{fig:decay} and Figure~\ref{fig:clear}. M. Wing acknowledges the support of the Alexander von Humboldt Stiftung and DESY, Hamburg. 



\bibliographystyle{elsarticle-num}
\bibliography{bibliography}


%
%
%
\end{document}